\documentclass[showpacs,twocolumn,aps,pre]{revtex4}
\usepackage{amssymb}

\usepackage{graphicx}
\usepackage{dcolumn}
\usepackage{bm}

\def\be{\begin{equation}}
\def\ee{\end{equation}}
\def\bea{\begin{eqnarray}}
\def\eea{\end{eqnarray}}
\def\a{\alpha}
\def\b{\beta}

\def\e{\epsilon}

\def\l{\lambda}

\begin{document}

\title{The Pfaffian solution of a dimer-monomer problem:\\
Single monomer on the boundary} \vskip 1cm
\author{F. Y. Wu}
\affiliation{Department of Physics,
 Northeastern University, Boston, Massachusetts 02115, U.S.A.}
 %\date{\today}

\begin{abstract} We consider the dimer-monomer problem for the rectangular
lattice. By mapping the problem into one of close-packed dimers on an extended
lattice, we rederive the Tzeng-Wu solution
for a single monomer on the boundary by evaluating a Pfaffian.
We also clarify the mathematical content of the Tzeng-Wu solution                              
by identifying it as the product of the nonzero eigenvalues of the Kasteleyn matrix.
  \end{abstract}


 \maketitle

\section{Introduction}
An outstanding unsolved problem in lattice statistics is the dimer-monomer
problem. While it is known \cite{lieb} that the
dimer-monomer system does not exhibit a phase transition, there
have been only limited closed-form results. 
The case of close-packed dimers on planar lattices has
been solved by Kasteleyn \cite{kas} and by Temperley and Fisher
\cite{ft,fisher}, and the solution has been extended to nonorientable
surfaces \cite{luwu,tesler}. But the general
dimer-monomer problem has proven to be intractable \cite{jerrum}.

\medskip
In 1974 Temperley \cite{temp} pointed out a bijection between
configurations of a single monomer on the boundary of a planar lattice 
and spanning trees on a related lattice. The bijection was 
used in  \cite{temp} to explain why enumerations of close-packed dimers
and spanning trees on square lattices yield the same
Catalan constant. More recently Tzeng and Wu \cite{tw} made
further use of the Temperley bijection to obtain the closed-form
generating function for  a single
monomer on the boundary. The derivation is however indirect since it
makes use of the Temperley bijection which obscures 
 the underlining mathematics of the closed-form solution.
 
\medskip
 Motivated by the Tzeng-Wu result, there has been renewed
interest in the general 
dimer-monomer problem. In a series of papers Kong \cite{kong,kong1,kong2}
has studied numerical enumerations of such
 configurations on $m\times n$ rectangular lattices for
varying $m,n$, and extracted finite-size correction terms for the single-monomer \cite{kong}
and general monomer-dimer \cite{kong1,kong2} problems. Of particular
interest is the finding \cite{kong} that in the case of a single monomer
the enumeration
exhibits a regular pattern similar to that found in the Kasteleyn solution
of close-packed dimers. This suggests that the general
single-monomer problem might be soluble. 

\medskip
 As a first step toward finding that solution
it is necessary to have 
an alternate and direct derivation of
the Tzeng-Wu solution without the recourse of the Temperley
bijection. Here we present such a derivation.
Our new approach points to the way of possible extension toward
the general single-monomer problem. It also identifies that,
besides an overall constant,
the Tzeng-Wu
  solution is  given by the 
square root of the product of the nonzero eigenvalues
of the Kasteleyn matrix, and thus clarifies its
underlining mathematics. 
  
\section{The single monomer problem}
Consider a rectangular lattice ${\cal L}$ consisting
of an array of $M$ rows and $N$ columns, where
both $M$ and $N$ are odd. The lattice consists of two
sublattices $A$ and $B$.  Since the total number of sites $MN$ is
odd, the four  corner sites belong  to the same sublattice, say, $A$
and there are one more $A$ than $B$ sites. The lattice can
therefore be completely covered by dimers if one $A$
site is left open. The open $A$ site can be regarded as a monomer.

\medskip
Assign non-negative weights $x$ and $y$ respectively to horizontal and
vertical dimers. When  the monomer
is on the
boundary, Tzeng and Wu \cite{tw}  obtained the following
closed-form expression for the generating function,
\bea
&& G(x,y)
= x^{(M-1)/2}y^{(N-1)/2} \nonumber \\
&& \times \prod_{m=1}^{\frac {M-1} 2}
\prod_{n=1}^{\frac {N-1} 2} \bigg[ 4 x^2 \cos^2 \frac {m\pi}{M+1}
+4 y^2 \cos^2 \frac {n\pi}{N+1} \bigg].\label{part}
 \eea
 This result
is independent of the location of the monomer provided that it is on 
the boundary.

\medskip
We  rederive the result (\ref{part}) using a formulation 
which is applicable to  any dimer-monomer problem. We  
first expand ${\cal L}$
into an extended lattice ${\cal L}'$ constructed by
connecting each  site occupied by a monomer to a new added site, and then consider 
close-packed dimers on ${\cal L}'$. Since the newly added sites are all
 of degree 1, all edges originating from the new sites must be covered by dimers.  
Consequently,   the  dimer-monomer problem
on ${\cal L}$ (with fixed monomer sites)
 is mapped to a close-packed dimer problem
on ${\cal L}'$, which can  be treated by standard means.

\medskip
We use the Kasteleyn method   \cite{kas} to treat the latter problem.
Returning to the single-monomer problem 
let the boundary monomer be at  site $s_0=(1,n)$
as demonstrated in Fig. 1a. The site $s_0$ is connected to a new site  $s'$ 
  by an edge
with weight $1$ as shown in Fig. 1b.
To enumerate close-packed dimers on ${\cal L}'$ using the Kasteleyn approach,
 we need to
orient, and associate phase factors to, edges so that all terms in
the resulting Pfaffian yield the same sign.

\begin{figure}
\center{\includegraphics [angle=0,height=1.5in]{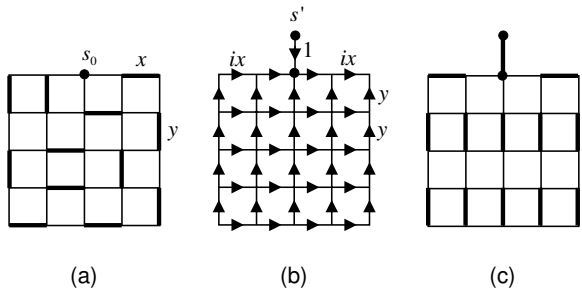}}
\caption{ (a) A dimer-monomer configuration on a $5\times 5$ lattice ${\cal L}$
with a single monomer at $s_0 =(1,3)$. 
(b) The extended lattice ${\cal L}'$ with edge orientation and a phase factor $i$
to horizontal edges.  (c) The reference dimer configuration $C_0$.}
\end{figure}

\medskip
 A convenient choice of orientation and assignment of  phase
factors is the one suggested by T. T. Wu \cite{ttwu}.  While Wu considered the case
$MN=$ even, the consideration can be extended to the present case.
Orient all horizontal
(resp. vertical) edges in the same direction 
 and  the new edge from $s'$ to $s_0$, and introduce a phase factor $i$
to all horizontal edges  as shown in Fig. 1b.  Then all
terms in the Pfaffian assume the same sign. To prove
this assertion it suffices to show that 
a typical term in the Pfaffian associated with a dimer configuration $C$ 
has the same sign as 
the term associated with a reference configuration $C_0$.
For $C_0$ we choose the configuration shown in Fig. 1c,
in which  horizontal dimers are placed in the first row  
with vertical dimers covering the rest of the lattice.
Then  $C$ and $C_0$ assume the same sign.

\medskip
The simplest way to verify the last statement is to
 start from a configuration in which every heavy edge in
$C_0$ shown in Fig. 1c is occupied by two dimers, and view
each of the doubly occupied dimers as a polygon of two edges.
 Then the `transposition
polygon' (Cf. \cite{kas}) formed by superimposing any $C$ and $C_0$ can always
be generated by deforming some of the doubly occupied edges into bigger polygons,
a process which does not alter the overall sign.  It follows that  $C$ and $C_0$ have the
same sign for any $C$.  This completes the proof.

\medskip
Here we have implicitly made use of the fact that the monomer is on the boundary.
If the monomer resides in the interior of ${\cal L}$, then there exist transposition polygons 
encircling the monomer site which may not necessarily carry the correct sign.
The Pfaffian, while can still  be evaluated,  does not yield 
the dimer-monomer generating function. We shall consider this general single-monomer
problem  subsequently \cite{luwu1}.

\medskip
With the edge orientation and phase factors in place,
the dimer generating function $G$ is  obtained by evaluating a Pfaffian 
\be
G(x,y) = {\rm Pf} (A')= \sqrt {{\rm Det}\ A'} \label{G}
\ee
where $A'$ is the antisymmetric Kasteleyn matrix of dimension $(MN+1)\times (MN+1)$. 
Explicitly, it reads
\be
A' = \pmatrix {0&0&\cdots&0&1&0&\cdots&0\cr
                0 & & & & & & & \cr
                \vdots&  & & &  & &  & \cr
                0&  & & &  & &  & \cr
                -1& & & &A & & &  \cr
                0 & &  & &  & & &  \cr
                \vdots&  & & &  & &  & \cr
                0 & & & &  & & &  \cr }. \label{A0matrix}
\ee
Here, $A$ is the  Kasteleyn  matrix of dimension $MN$
for ${\cal L}$ given by
\be
A = i\,x\,T_M \otimes I_N + y\, I_M \otimes T_N, \label{Amatrix}
\ee
with $I_N$  the $N\times N$ identity matrix and $T_N$  the $N\times N$ matrix
\be
T_N = \pmatrix{ 0& 1 & 0& \cdots & 0 &0 \cr
               -1& 0 & 1& \cdots & 0 &0 \cr
                0& -1& 0& \cdots & 0 &0 \cr
           \dots& \dots &\dots &\dots&\dots&\dots \cr
                0& 0& 0& \dots & 0&1 \cr
                0& 0& 0& \dots & -1&0}.  \label{Tmatrix}
\ee
Note that elements of $A$ are labeled by $\{(m,n);(m',n')\}$,
where $(m,n)$ is a site index, and  the 
element $1$ in the first row of $A'$ is at position $(1,n)$ of $A$, $n=$ odd.

\medskip
Expanding   
(\ref{A0matrix}) along the first row and column, we obtain   
\be
{\rm Det}\, A' = C(A;\{( 1,n);(1,n)\}) \label{Aprime}
\ee
where $C(A;\{( 1,n);(1,n)\}) $
 is the cofactor
of the $\{(1,n);(1,n)\}$-th  element of ${\rm Det}\, A$.

\medskip
The cofactor $C(\a,\b)$ of the $(\a,\b)$-th element of any 
non-singular $A$ can be computed using the identity 
\be
C(A; \a,\b) = A^{-1}(\b,\a) \times {\rm Det} \, A, \label{cofactor}
\ee
where $A^{-1}(\b,\a)$ is the $(\b,\a)$-th element of $A$.
However,  the formula is not directly
useful in the present case since the matrix $A$ is singular. 
We shall return to its evaluation  in Sec. IV.

\section{Eigenvalues  of the determinant $A$}
In this section we enumerate  the eigenvalues of $A$. 

\medskip
 The matrix
 $T_N$ is diagonalized by the similarity transformation
\bea
U^{-1}_N \, T_N \,  U_N = \Lambda _N \nonumber
\eea
where $U_N$ and $U^{-1}_N$ are $N\times N$ matrices with elements
\bea
U_{N}(n_1,n_2) &=& \sqrt \frac 2 {N+1}\,i\,^{n_1} \sin \frac {n_1n_2\pi}{n+1} \nonumber \\
U_{N}^{-1} (n_1,n_2) &=& \sqrt \frac 2 {N+1} \, (-i)^{n_2}  \sin \frac {n_1n_2\pi}{N+1} , \label{UU}
\eea
and $\Lambda_N$ is an $N\times N$ diagonal matrix whose diagonal elements are
the eigenvalues of $T_N$,
\be
\lambda_m = 2i\, \cos   \frac {m\pi}{N+1}, \quad m = 1,2,\cdots, N. 
\ee

\medskip
Similarly the $MN \times MN$ matrix $A$ is diagonalized by the similarity
 transformation generated by $U_{MN} = U_M\otimes U_N$.  Namely,
\bea
U_{MN}^{-1} \,A\, U_{MN} = \Lambda_{MN} \label{UMN}
\eea
where $\Lambda_{MN}$ is a diagonal matrix with eigenvalues
\bea
\l_{mn} &=& 2 i \bigg[i\, x \,\cos   \frac {m\pi}{M+1} + y\, \cos
  \frac {n\pi}{N+1}\bigg], \nonumber \\
 && \hskip .5cm m=1,2,...,M,\, n=1,2,...,N,
 \label{eigenvalue}
\eea
on the diagonal, and elements of $U_{MN}$ and $U^{-1}_{MN}$ are
\bea
U_{MN}(m_1,n_1;m_2,n_2) &=& U_M(m_1,m_2)\, U_N(n_1,n_2) \nonumber \\
U^{-1}_{MN}(m_1,n_1;m_2,n_2) &=& U^{-1}_M(m_1,m_2) \, U^{-1}_N(n_1,n_2) .\nonumber
\eea
Then we have
 \be
{\rm Det}\, A = \prod_{m=1}^M \prod_{n=1}^N \l_{mn}. \label{DetA}
\ee

As in (\ref{G}), close-packed dimers on ${\cal L}$ are enumerated by 
evaluating $\sqrt { {\rm Det} A}$.
For $MN$  even, this procedure gives precisely the Kasteleyn solution  \cite{kas}.
 For $MN$  odd, the case we are considering, the eigenvalue
$\l_{mn} = 0$ for $m=(M+1)/2, n=(N+1)/2$, and hence
 ${\rm Det}\, A = 0$, indicating correctly  there is no dimer covering 
of ${\cal L}$.
However, it is useful for later purposes  to consider
the product of the nonzero
eigenvalues of $A$,
\be
P\equiv {{\prod_{m=1}^M \prod_{n=1}^N}}\, '\, \l_{mn} , \label{P1}
\ee
where the prime over the product denotes the restriction
$ (m,n) \not= \big(\frac {M+1} 2,\frac {N+1} 2 \big)$.

\medskip 
Using the identity
\bea
\cos \bigg( \frac {m}{M+1}\bigg) \pi = - \cos  \bigg( \frac {M-m+1}{M+1}\bigg) \pi, \nonumber
\eea
one can rearrange factors in the product to arrive at
\be
P = Q \,\prod_{m=1}^{\frac {M-1} 2}  \prod_{n=1}^{\frac {N-1} 2} \, 
\bigg( 4 x^2 \cos^2 \frac {m\pi}{M+1} +4 y^2 \cos^2 \frac {n\pi}{N+1}\bigg)^2 
\label{P}
\ee
where the factor $Q$ is the product of factors with either $m=\frac {M+1} 2$
or $n=\frac {N+1} 2$.  Namely,
\bea
Q &=&  \Bigg[ \prod_{m=1}^{\frac {M-1} 2}  4x^2\cos ^2 \frac {m\pi}{M+1} \Bigg] \times
      \Bigg[ \prod_{n=1}^{\frac {N-1} 2}  4y^2\cos ^2 \frac {n\pi}{N+1} \Bigg] \nonumber \\
   &=& \bigg[\frac {(M+1)(N+1)}4\bigg] x^{M-1}  y^{N-1}, \label{Q}
\eea
where we have made use of the identity
\bea
\prod_{n=1}^{\frac {N-1} 2}\bigg( 4 \cos^2 \frac {n\pi}{N+1}\bigg) = \frac {N+1} 2,
\quad N= {\rm odd}. \nonumber
\eea
The expression (\ref{P}) for $P$ will be used in the next section.

\section{Evaluation of the cofactor}
We now return to  the evaluation of the cofactor 
$C(A;\{( 1,n);(1,n)\}) $. 
We shall however evaluate the cofactor $C(A;\{( m,n);(m',n')\}) $
for  general $m,m',n,n'$,
 although only the result of $m=m'=1,\, n =n'$ is needed here.

\medskip
 To circumvent the  problem of using (\ref{cofactor}) caused by 
the vanishing of    ${\rm Det}\, A =0$, we
  replace $A$  by the matrix
\bea
A(\e) =  A + \e \, I_{MN}\, , \quad \e  \not= 0 \nonumber 
\eea
whose inverse exists, and take the $\e \to 0$ limit to rewrite
(\ref{cofactor}) as
\bea
&& C(A;\{( m,n);(m',n')\}) \nonumber \\
 && \quad =\lim _{\e \to 0} \bigg[  \big[A^{-1}(\e)\big] (m',n';m,n) \times
 {\rm Det}\ A(\e) \bigg] .\label{part1}
\eea
Quantities on the r.h.s. of (\ref{part1}) are now well-defined and the cofactor 
can be evaluated accordingly. 
The consideration of the inverse
of a singular matrix along this line is known in mathematics literature as
finding the pseudo-inverse \cite{inverse,inverse1}. The method 
of taking the small $\epsilon$ limit used here
has previously been used successfully in the analyses of resistance \cite{resistor}
and impedance \cite{impedance} networks.
 
\medskip
The eigenvalues of $A(\e) $ are $\l_{mn}(\e) =\l_{mn} + \e$ and hence we have
 \be
{\rm Det}\, A(\e) = \prod _{m=1}^M \prod_{n=1}^N \big[\l_{mn} +\e \big]
= \e\, P + O(\e^2),\label{Aepsilon}
\ee
where $P$ is the product of nonzero eigenvalues given by (\ref{P}).

\medskip
We next evaluate $A^{-1}(\e)(m.n;m,n)$
and retain only terms of the order of $1/\e$.  Taking the inverse of (\ref{UMN})
with $A(\e)$ in place of $A$, we obtain
\bea
A^{-1}(\e) = U_{MN}\, \Lambda_{MN}^{-1}(\e)\, U_{MN}^{-1}. \nonumber
\eea
Writing out its   matrix elements explicitly, we have
\begin{widetext}
\be
A^{-1}(\e) (m',n'; m,n) = \sum_{m''=1}^M
\sum_{n''=1}^N \frac {U_{MN}(m',n';m'',n'') U^{-1}_{MN}(m'',n'';m,n)} {\l_{m'',n''}+\e}.
\ee
For $\e$ small  the leading term comes from $(m'',n'')= (\frac {M+1} 2, \frac {N+1} 2)$ 
for which $\l_{m'',n''} =0$.  Using $U^{-1}_{MN}(m,n;m',n') = U^{-1}_{M}(m,m')\,U^{-1}_{N}(n,n')$ 
and (\ref{UU}), 
this leads to the expression
\bea
A^{-1}(\e) (m',n'; m,n) &=& \bigg(\frac 1 \e\bigg)
\bigg[\frac {4\, i^{m'+n'} (-i)^{m+n} } {(M+1)(N+1)} \bigg] \sin  \frac {m'\pi} 2 
\sin  \frac {n'\pi} 2 \sin  \frac {m\pi} 2 \sin  \frac {n\pi} 2 +\  O(1). \nonumber
\eea
 \end{widetext}
Thus, after making use of  ({\ref{part1}) and (\ref{Aepsilon})  we obtain
\bea
&&C(A;\{( m,n);(m',n')\}) \nonumber \\
 && =\sin \frac {m \pi} 2 \sin \frac {n \pi} 2 
    \sin \frac {m' \pi} 2 \sin \frac {n '\pi} 2 
\bigg[  {\frac {4\, i^{m'+n'} (-i)^{m+n} P} {(M+1)(N+1)} }\bigg] .\nonumber \\
 \label{CA}
\eea

Finally, specializing to $m=m'=1, \,n=n'$ and combining (\ref{G}), (\ref{Aprime}),  and (\ref{CA}),
 we obtain
 \bea
G(x,y) &=& \sqrt{ C(A;\{( 1,n);(1,n)\}) } \nonumber \\
     &=&  \sqrt{  \frac {4 P} {(M+1)(N+1)}  }\, ,\hskip 0.2cm
 {\rm for \>\>} n {\rm \>\> odd\>\>} (A \>\> {\rm site}) \nonumber \\
   &=&  0, \hskip 2.9cm {\rm for \>\>} n {\rm \>\> even\>\>} (B  \>{\rm \>site})\nonumber \\
\label{final}
\eea
This gives the result (\ref{part}) after introducing (\ref{P}) for $P$. 
It also says that there is no dimer covering if the monomer is on a $B$ site.

\medskip
The expression
(\ref{final}) clarifies the underlining mathematical content of the 
Tzeng-Wu solution (\ref{part}) by identifying it as the product of the {\it nonzero} 
eigenvalues of the Kasteleyn matrix. 
This is compared to the Kasteleyn result \cite{kas} that
for $MN=$ even the dimer generating function is given by 
 the product of {\it all} eigenvalues.

\section{Discussions}
We have used a direct approach to derive the closed-form expression of the dimer-monomer
generating function for the rectangular lattice with
a single monomer on the boundary.   
Our approach is to first convert the
problem into one of close-packed dimers without monomers,
and consider the latter problem using established means.
  This approach suggests a possible route toward analyzing
 the  general dimer-monomer problem.

\medskip
 We have also established that the Tzeng-Wu solution
(\ref{part}) is given by the product of the nonzero eigenvalues of
the Kasteleyn matrix of the lattice.   This is reminiscent to the
well-known result in algebraic graph theory \cite{graph} that
spanning trees on a graph are enumerated by evaluating the product
of the nonzero eigenvalues of its tree matrix. The method of
evaluating cofactors of a singular matrix as indicated by
(\ref{part1}), when applied to the tree matrix 
of spanning trees
details of which can be easily worked out, 
offers a simple and
direct proof of the fact that all cofactors of a tree matrix are
equal and equal to the product of its nonzero eigenvalues.
 The intriguing
similarity of the results suggests there might be something deeper
lurking behind our
 analysis.

\medskip
I am grateful to W. T. Lu for help in the preparation of the manuscript.

\end{document}